\documentclass[11pt,a4paper]{article}
\usepackage[hyperref]{acl2020}
\usepackage{times}
\usepackage{latexsym}
\usepackage{graphicx}

\usepackage{microtype}

\aclfinalcopy 

\title{A Hybrid BERT and LightGBM based Model for Predicting Emotion GIF Categories on Twitter}

\author{Ye Bi \\
  Ping An Technology(Shenzhen) \\
  Shanghai, China \\
  \texttt{magicyebi@163.com} \\\And
  Shuo Wang \\
  Ping An Technology(Shenzhen) \\
  Shanghai, China \\
  \texttt{wangshuotzjz@163.com} \\\And
  Zhongrui Fan \\
  Ping An Technology(Shenzhen) \\
  Shanghai, China \\
  \texttt{wo459847749@163.com} \\}

\date{}

\begin{document}
\maketitle
\begin{abstract}
The animated Graphical Interchange Format~(GIF) images have been widely used on social media as an intuitive way of expression emotion.   
Given their expressiveness, GIFs offer a more nuanced and precise way to convey emotions.
In this paper, we present our solution for the EmotionGIF 2020 challenge, the shared task of SocialNLP 2020.
To recommend GIF categories for unlabeled tweets, we regarded this problem as a kind of matching tasks and proposed a learning to rank framework based on Bidirectional Encoder Representations from Transformer~(BERT) and LightGBM.
Our team won the 4th place with a Mean Average Precision @ 6 (MAP@6) score of 0.5394 on the round 1 leaderboard. 
\end{abstract}

\section{Introduction}

Animated GIFs are widely used in online social media nowadays. 
They provide a way to convey emotions more accurately and conveniently. But algorithms on those GIFs are still insufficient. 
Fortunately, EmotionGIF 2020 \cite{shmueli2020socialnlp} gives us a chance to have a glimpse of the relation between animated GIFs predicting and emotion detection. 
The task of EmotionGIF is defined as given unlabeled tweets, predict the category of a GIF response. The evaluation metric is Mean Average Precision at 6(MAP@6).

Unlike traditional emotion detection, a single animated GIF may contain various emotions, also may contain different set of emotions under different circumstances, making it harder to predict.
After referring to several classical Natural Language Processing(NLP) models like variations of Transformers \cite{DBLP:journals/corr/VaswaniSPUJGKP17} and BERT \cite{DBLP:journals/corr/abs-1810-04805},
we find it may be unsuitable for those models to directly apply on the challenge task. 
Considering the multi-label of raw GIFs and the metric, we make improvement on BERT, then propose our own solution: a cascade pairwise classifier by using BERT and LightGBM together.

The rest of the paper is organized as follows: Section 2 describes the dataset and preprocessing techniques.
Our solution is introduced in Section 3.  
We show the experiments and results of our model in the next Section 4.  
Finally, we conclude our analysis of the challenge. 

\section{Dataset and preprocessing}
\subsection{Dataset}
The organizer builds a dataset from 40,000 two-turn twitter threads. 
The dataset is splitted into three parts: 32,000 samples with gold labels for training, 4000 unlabeled samples for practice and 4000 unlabeled samples for final evaluation.
Each sample includes the text of the original tweet and the reply text.
The training samples also contains the MP4-format file and the category of the GIF response, which is selected from a list of 43 categories.  
Any GIF response can belong to multiple categories and the label in the dataset is a non-empty subset of 1 to 6 categories.
The data in a sample is shown in Table~\ref{sample}.

\begin{table}
\centering
\begin{tabular}{ll}
\hline
\textbf{idx} & 32 \\
\textbf{text} & Fell right under my trap \\
\textbf{reply} & Ouch! \\
\textbf{mp4}  & fe6e...ff82.mp4 \\
\textbf{categories} & ['awww','yes','oops'] \\
\hline
\end{tabular}
\caption{\label{sample} A data sample.}
\end{table}

\subsection{Preprocessing}
The text preprocessing is a crucial step in many NLP tasks.
There are some research \cite{Magliani2016A} and tools on the text preprocessing of twitter.
We utilized NLTK’s TwitterTokenizer \cite{Steven2006NLTK} and performed some basic preprocessing steps on the tweet text:
\begin{itemize}
    \item convert all tweet texts into lower case.
    \item convert the emoji symbols to their corresponding meanings using emoji \footnote{https://pypi.org/project/emoji/} python package.
    \item remove excessive spaces. 
    \item user names, numbers and websites are replaced with special tokens. 
\end{itemize}

\section{Methodology}
\subsection{Model Overview}

\begin{figure*}[!htbp]
\centering
\makeatletter
{\includegraphics[width=7in]{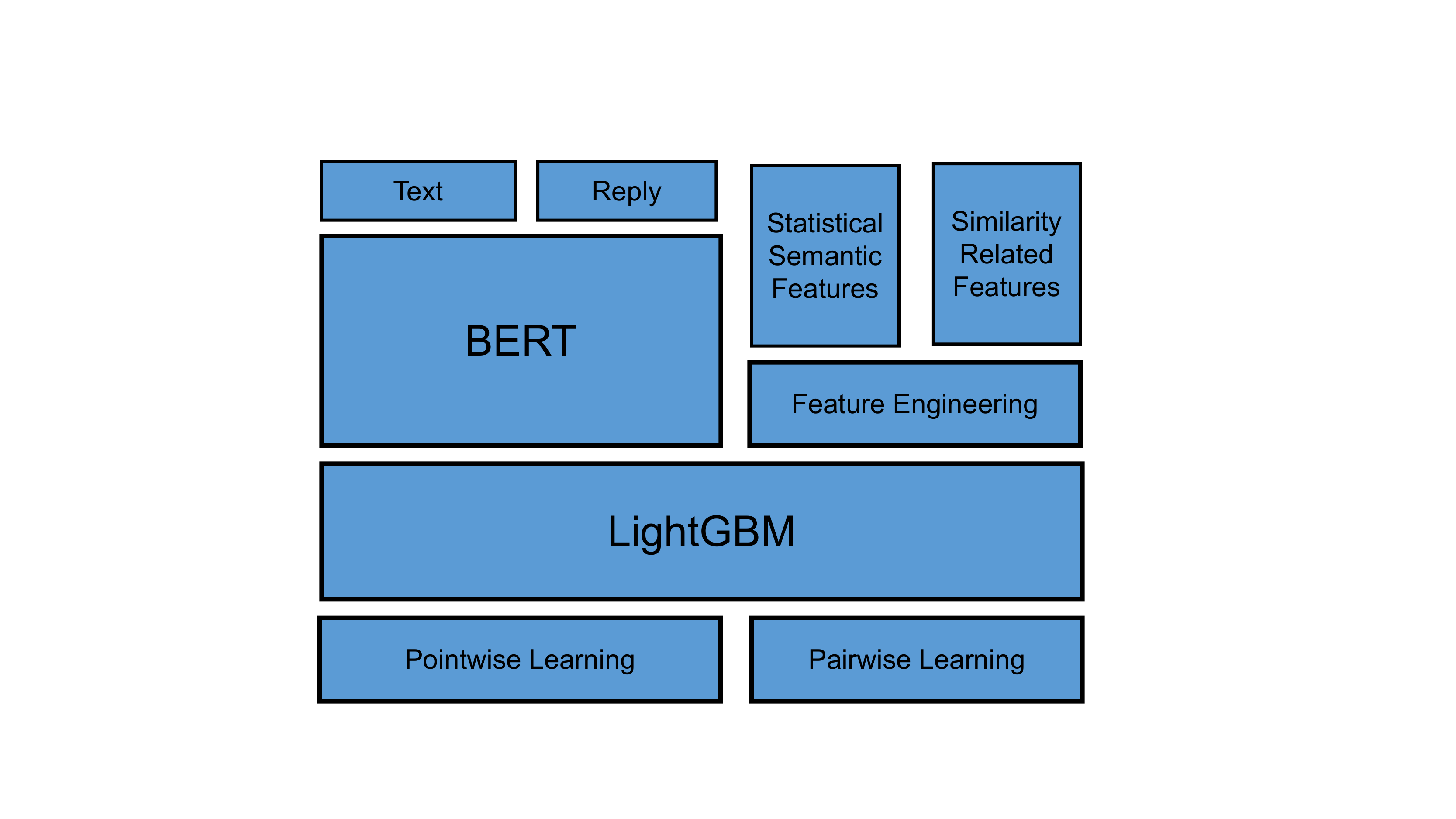}}
\makeatother
\caption{The Model Overview.}
\label{overview}
\end{figure*}

In this section, we first introduce the BERT model with pointwise and pairwise approaches, respectively.
And then we describe the LightGBM model in details.
The pipeline of the proposed method is shown in Figure~\ref{overview}. 

\subsection{BERT Model}
BERT model has achieved great success in many NLP tasks such as sentiment classification, natural language inference, and machine reading comprehension. 
So it's a natural choice to employ BERT in our framework.
We fine-tune the BERT model in two different ways: pointwise scheme and pairwise scheme.
\subsubsection{BERT with pointwise}
We trained the BERT model in the pointwise way, which means that we treated the task as the binary classification problem.
We concatenated the preprocessed tweet text and reply with [SEP] token and sent them into the BERT model as text pair.
The pooled output of BERT model is treated as the representation of tweet text. 
The 43 GIF categories are embedded into a high dimensional space. 
A score function is defined to measure how well the tweet content and the GIF category match.
Since the given labeled samples are all positive samples, we need to match some negative samples during the training process.

For pointwise approach, the score for each category is independent of the other categories that are in the result list for the query.
That is to say, the pointwise method can't take into account the internal dependencies between the categories corresponding to the same query.
This leads us to take pairwise model into consideration.

\subsubsection{BERT with pairwise}

Pairwise approaches look at a pair of categories at a time in the loss function. 
Given a pair of categories, they try to come up with the optimal ordering for that pair.
To be specific, each training sample consists of tweet text query and two GIF categories, one is the true matched category, the other is the negative category sampled randomly from categories excluding the true category. 
The pairwise model learns to compare the differences between the query and each category and pick the true one. 
We used the margin ranking loss as our loss function and trained several triplet samples with the same tweet text and different GIF categories. 
We got a higher score than the BERT model in pointwise mode.

\subsection{LightGBM Model}

We use pairwise-way to train this classifier. Its inputs contains multiple features, including the embedding returned by the pointwise/pairwise BERT and other manual features. 
And the output is the score upon the specified label given.
We mainly create similarity scores between sentence(text, reply) and label as manual features due to the pairwise-way training. Different NLP algorithms are used to generate similarity vectors, for instance, TF-IDF transformation,
Word2Vec pretraining, FastText pretraining, etc.. In order to gain different fine-grained language, we build the corpus in two ways. For word-level granularity, we separate all words in all sentences after removing the emoji in it.
For sentence-level granularity, we directly use the sentence after removing the emoji. We choose three algorithms to calculate the similarity scores, including Euclidean distance, Manhattan distance and Cosine distance. 
For labels, the similarity vectors is incalculable actually, so we aggregate the vectors belong to a specified label by using mean pooling as the vector of this specified label. 
The same method is used again upon the embedding returned by the pointwise/pairwise BERT. Besides, we also create some statistical features such as the number of emoji, the weight of key words, etc.



\section{Experiment}
\subsection{Experimental Settings}

In our experiments, the 32000 training samples are splitted into training set and validation set at a ratio of 9:1.
In the training phase, each positive sample is paired with 4 negative samples.
For the BERT-base model, The embedding dimension of GIF category is set to 128.
The code is implemented under \emph{PyTorch} framework and is trained with \emph{Adam} optimizer.
The initial learning rate is 0.00003 and decays every 10 epochs at a rate of 0.1.
The training of BERT-base model takes about 4 hours on 4 NVIDIA Tesla V100 GPUs for 30 epochs with a minibatch size of 128.

For the LightGBM-base model, we use hyperopt \footnote{https://github.com/hyperopt/hyperopt} for hyperparameter optimization.

\subsection{Results}

Here we compare the performance of our method with different settings.
The results on the validation set are shown in Table~\ref{results}. 
From the Table, we can see that the result of the pairwise training method is better than pointwise. 
At the same time, the LightGBM model based on detailed feature engineering is very effective.

\begin{table}
\centering
\begin{tabular}{lcc}
\hline \textbf{Model} & \textbf{Offline} & \textbf{Online}\\ \hline
BERT(pointwise) & 0.5172 & - \\
BERT(pairwise) & 0.5354 & 0.5209 \\
LightGBM & 0.5645 & 0.5394 \\
\hline
\end{tabular}
\caption{\label{results} MAP@6 results of different settings.}
\end{table}

\section{Conclusion}
In this paper, we propose a method based on BERT and LightGBM under pairwise learning framework for EmotionGIF 2020 Challenge. 
Extensive experiments were conducted on the challenge dataset and the results proved the effectiveness of our method.

\bibliography{anthology,acl2020}
\bibliographystyle{acl_natbib}

\appendix

\end{document}